\documentclass[
12pt,
superscriptaddress,
showpacs, showkeys,
amsmath,amssymb,
aps,
pre,
floatfix,
]{revtex4-1}

\usepackage{hyperref}
\usepackage{graphicx}
\usepackage{amsmath}
\usepackage{amsfonts}
\usepackage{amssymb}
\usepackage{yfonts}
\usepackage{bm}
\usepackage{siunitx}
\usepackage{enumitem}
\usepackage{color}

\begin{document}

\title{Density-Dependent Finite System-Size Effects in Equilibrium Molecular Dynamics Estimation of Shear Viscosity: Hydrodynamic and Configurational Study}

\author{Kang-Sahn Kim}
\affiliation{Department of Chemistry, Korea Advanced Institute of Science and Technology (KAIST), \\ Daejeon 34141, Republic of Korea}

\author{Changho Kim}
\email{ckim103@ucmerced.edu}
\affiliation{Department of Applied Mathematics, University of California, Merced, California 95343, USA}

\author{George Em Karniadakis}
\affiliation{Division of Applied Mathematics, Brown University, \\ Providence, Rhode Island 02912, USA}

\author{Eok Kyun Lee}
\affiliation{Department of Chemistry, Korea Advanced Institute of Science and Technology (KAIST), \\ Daejeon 34141, Republic of Korea}

\author{John J.\ Kozak}
\affiliation{Department of Chemistry, DePaul University, Chicago, Illinois 60604, USA}

\date{\today}

\begin{abstract}
We study the intrinsic nature of the finite system-size effect in estimating shear viscosity of dilute and dense fluids within the framework of the Green--Kubo approach.
From extensive molecular dynamics simulations, we observe that the size effect on shear viscosity is characterized by an oscillatory behavior with respect to system size $L$ at high density and by a scaling behavior with an $L^{-1}$ correction term at low density.
Analysis of the potential contribution in the shear-stress autocorrelation function reveals that the former is configurational and is attributed to the inaccurate description of the long-range spatial correlations in finite systems.
Observation of the long-time inverse-power decay in the kinetic contribution confirms its hydrodynamic nature.
The $L^{-1}$ correction term of shear viscosity is explained by the sensitive change in the long-time tail obtained from a finite system.
\end{abstract}

\maketitle{}

\section{Introduction}

The shear viscosity coefficient in dense fluids, along with diffusion coefficient and thermal conductivity, constitute important ingredients of hydrodynamic theory.
These transport coefficients are connected to the corresponding time-correlation functions of microscopic fluctuating variables via the Green--Kubo relations.
At equilibrium, or in small deviations from equilibrium, a systematic connection between the correlation functions and the hydrodynamic equations has been established in the long-wavelength and small-frequency hydrodynamic limit~\cite{HansenMcDonald2013, BoonYip1980, BalucaniZoppi1995, Forster1975}.

The molecular dynamics (MD) simulation technique provides several methods to evaluate the shear viscosity coefficient of a molecular fluid.
First, the shear viscosity coefficient can be estimated by comparing flow patterns generated from nonequilibrium MD simulations with those predicted by hydrodynamic theory~\cite{Erpenbeck1984, EvansMorriss1989, BackerLowe2005}.
While these direct methods can, in principle, be used to calculate nonlinear as well as linear transport coefficients, they have an inherent arbitrariness in producing the nonequilibrium flow fields with respect to the thermostat, barostat, and other factors that influence the motion of individual particles by an external applied force.
The second type of method examines the long-time behavior of the long-wavelength correlations in the equilibrium fluctuations of the transverse momentum field~\cite{Palmer1994}.
It is based on the observation that in the hydrodynamic limit, these correlations decay exponentially with the exponent proportional to the shear viscosity.
The third approach is to compute the shear viscosity $\eta$ using the Green--Kubo integral~\cite{LevesqueVerletKurkijarvi1973, Erpenbeck1988, MaginnMesserlyCarlsonRoeElliott2018},
\begin{equation}
\label{GK_viscosity}
\eta = \int_0^\infty C(t)dt,\quad C(t)=\frac{V}{k_\mathrm{B}T}\langle p_{xy}(0)p_{xy}(t)\rangle.
\end{equation}
Here, $C(t)$ is the normalized shear-stress autocorrelation function (SACF), which is calculated from an equilibrium MD simulation of a system with volume $V$ at temperature $T$.
Contrary to the other two approaches, detailed information about the dynamics of the microscopic fluctuating variable (i.e., the off-diagonal pressure tensor component $p_{xy}$) is available.

\subsection{Density Dependence}

Structural relaxation in dense fluids is a complex process and coupled with momentum relaxation.
Since the microscopic expression of $p_{xy}$ [See Section~\ref{subsec_decomp_sacf}] is written as the sum of kinetic (K) and potential (P) contributions,
\begin{equation}
\label{decomp_p}
p_{xy} = p_{xy}^\mathrm{K} + p_{xy}^\mathrm{P},
\end{equation}
these relaxation processes can be investigated from the resulting decomposition of the SACF into kinetic-kinetic (KK), potential-potential (PP), and kinetic-potential (KP) components, 
\begin{equation}
\label{deomp_SACF}
C(t) = C^\mathrm{KK}(t) + C^\mathrm{PP}(t) + 2 C^\mathrm{KP}(t).
\end{equation}
While kinetic theory has been successful in explaining the behavior of kinetic term~\cite{ErnstHauge1971, ErpenbeckWood1982}, it has not given a complete answer for theoretical substantiation of the asymptotic long-time behavior of the SACF and prediction of density dependence of the shear viscosity.
This is because the behaviors of the potential and cross terms exhibit distinctively different features from the kinetic term.
In particular, in the high density region near the fluid-solid transition point, high-frequency viscoelastic modes play an important role but their influence on the shear viscosity is still not clearly understood.

\begin{figure}
\includegraphics[angle=0, width=0.95\textwidth]{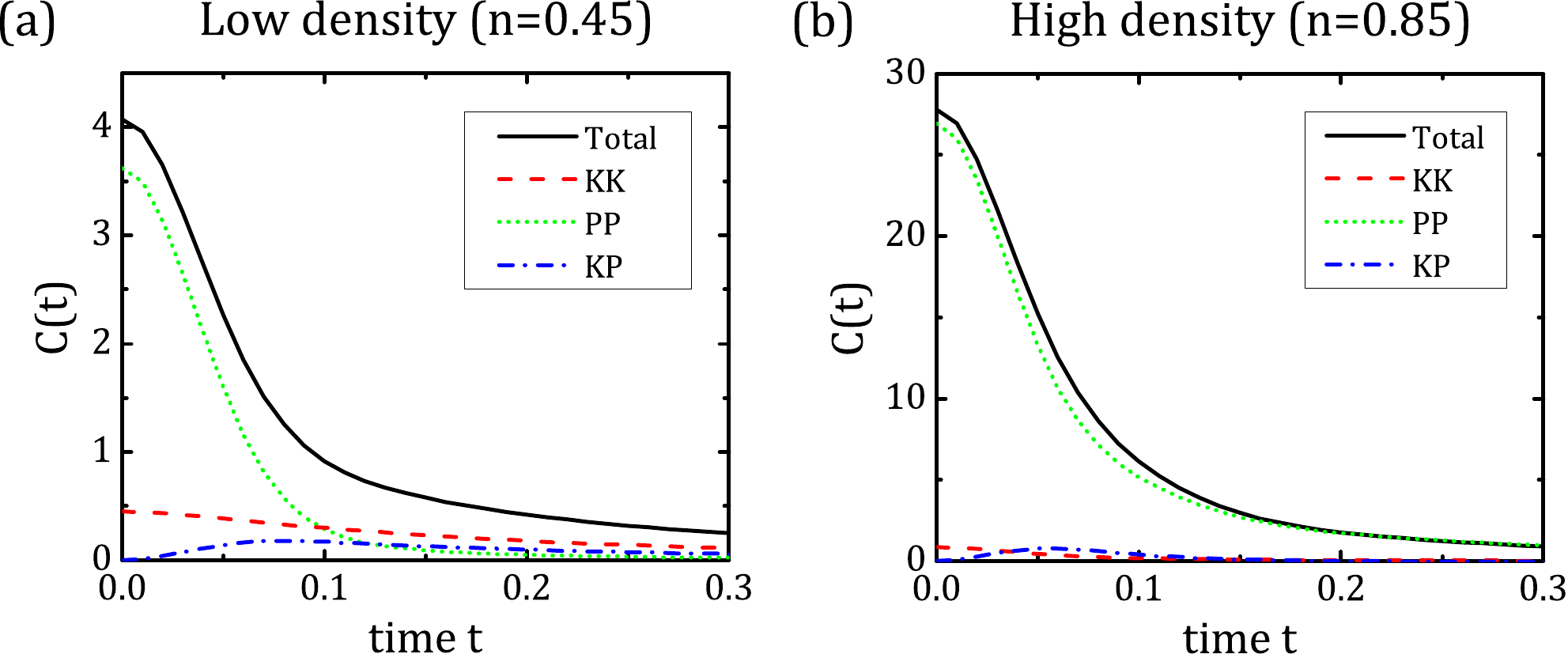}
\caption{\label{fig:sep_sacf}
Decomposition of the normalized SACF $C(t)$ (defined in Eq.~\eqref{GK_viscosity}) into three components: potential-potential (PP), kinetic-kinetic (KK), and kinetic-potential (KP).
For a simple fluid interacting via the WCA potential, results from two number densities, $n=0.45$ and 0.85, are shown in panels~(a) and (b), respectively.
Corresponding log-log plots are shown in Figure~\ref{fig:sep_sacf_ln}. 
Simulation details are provided in Section~\ref{sec_methods}.}
\end{figure}

As illustrated in Figure~\ref{fig:sep_sacf}, the SACF at high density is dominated by the potential term, whereas the long-time behavior of the SACF at low density is governed by the kinetic term.  
A significant contribution by the potential term explains why the prediction of the shear viscosity using kinetic theory fails in the high density region.
At low densities, a substantial increase in the kinetic contribution implies that the dynamics is no longer dominated by structural relaxation.
Competition between the two contributions, in addition to that of the cross term, makes the physical nature of the density dependence of the shear viscosity difficult to understand.

Previous MD simulation results~\cite{IsobeAlder2009, IsobeAlder2010, IsobeAlder2012} suggest an intimate connection between shear viscosity and its configurational nature at high density.
The configurational nature includes aspects of the dependence of the shear stress on the atomic configuration.
For example, the long-time tail of the SACF observed in high density region, often referred to as the ``molasses tail,'' has a fundamentally different physical origin from the hydrodynamic vortex flow of the velocity autocorrelation function~\cite{ChoiHan2017, HanKimTalknerKarniadakisLee2018} and also from the kinetic fluxes related to shear viscosity~\cite{ErnstHauge1971}.
This molasses tail effect becomes more pronounced as fluid density approaches the fluid-solid transition point and results in a markedly enhanced shear viscosity near solidification.
MD simulation results on the molasses tail reveals that it is due to the orientational correlations of bonds connecting colliding pairs of particles at long times and high densities near solidification.
The existence of algebraic tails, followed by the stretched exponential, is understood as having a hydrodynamic origin which includes much shorter wavelength fluctuations~\cite{LaddAlder1989,LeegwaterBeijeren1989}.
The long-range many-body `excluded volume' effects showing up in the potential component $C^\mathrm{PP}(t)$ at high density have been noted for the Lennard-Jones fluid in the past~\cite{StassenSteele1995a, StassenSteele1995b} and an empirical equation of state of shear viscosity that is accurate in a broad range of density and temperature has been proposed~\cite{Woodcock2005}.

\subsection{Finite System-Size Effects}

In the process of analyzing the physical meaning of the simulation results, the effects of finite simulation system size should be effectively removed.
These effects are unavoidable and cause serious difficulty in the interpretation of the simulation results.
The periodic boundary conditions conventionally employed in MD simulations as an effort to correct physical properties observed in finite systems, and mimic the infinite system, still differ from material properties of the bulk system due to neglect of long-wavelength fluctuations of corresponding dynamical variables.

Compared to the self-diffusion coefficient, for which there have been a significant number of MD studies on the finite system-size effects since the celebrated Yeh--Hummer system-size correction~\cite{YehHummer2004}, there have been only a few studies that investigate the particular issue for the shear viscosity.
Most of the recent MD studies reported a weak system-size dependence~\cite{DaivisEvans1995, YehHummer2004, MoultosZhangTsimpanogiannisEconomouMaginn2016}.
However, it has not been clearly confirmed whether this system-size effect exhibits a scaling law.
Moreover, very little has been known about its density dependence, not to mention its physical interpretation.

Our previous work~\cite{KimHan2018} showed how configurational degrees of freedom influence values of the shear viscosity as the simulation system size changes at high density.
The MD simulation results for simple and complex fluids have identified the presence of system size-dependent behaviors of the shear viscosity at high densities, especially for the case of small system sizes owing to limited configurational rearrangements in finite systems~\cite{Petravic2004a, Petravic2004b, KimHan2018}.
This restriction leads to complex oscillatory behavior of the shear viscosity (in the plot of $\eta_L$ versus the system size $L$), which however quickly disappears for system sizes larger than a critical size.

It is, however, expected that the effect of finite system size in estimating the shear viscosity will have a density dependence if the potential and kinetic components of the SACF exhibit different system-size effects and compete with each other.
The behavior of the kinetic part has been theoretically studied using the mode-coupling theory with the linearized Navier--Stokes equations, and finite system-size effects on the algebraic tail of $C^\mathrm{KK}(t)$ have been investigated~\cite{ErnstHauge1971, ErpenbeckWood1982}.
The simulation system-size dependence of the SACF and the shear viscosity makes it even harder to understand the nature of their density-dependent behavior.
Hence, it is crucial to understand the density-dependent behavior and relative weight of each contribution to the SACF to discern the actual behavior of the SACF and shear viscosity in the thermodynamic limit.

\vspace{2em}

In this paper, we investigate the behavior of the three contributors to the SACFs.
Our study focuses on the shear viscosity of simple fluids interacting via the Weeks--Chandler--Andersen (WCA) potential at three typical densities as a function of system size.
The SACFs obtained from equilibrium MD simulation exhibit a variety of different patterns depending on system density.
We relate system-size effects with the hydrodynamic behavior (induced by the kinetic component) as well as the configurational restriction (induced by the potential component).
It is noted that a comprehensive understanding of the behavior of the SACF has, until now, been a demanding task because the short- and long-time behavior of the SACF exhibit qualitative differences, and moreover, the long-time behavior sensitively depends on simulation system size.
The knowledge acquired from this study is theoretically important in understanding the hydrodynamic and configurational features in collective dynamics.
Our study also provides practical suggestions on how to evaluate the shear viscosity in thermodynamic limit.

The rest of the present paper is organized as follows.
In Section~\ref{sec_methods}, we present a brief description of our MD simulation as well as the decomposition of the SACF and the shear viscosity.
In Section~\ref{sec_res_sacf}, we discuss the density-dependent behavior of the SACF by examining the physical origins of characteristic long-time behaviors of the kinetic and potential components. 
In Section~\ref{sec_res_vis}, we investigate two types of finite system-size effect on the shear viscosity coefficient, which have hydrodynamic and configurational origins. 
Our conclusions are presented in Section~\ref{sec_conclusion}.

\section{\label{sec_methods}Method}

\subsection{\label{subsec_decomp_sacf}Decomposition of SACF}

For a simple fluid system under periodic boundary conditions, the molecular expression of the off-diagonal pressure tensor component $p_{xy}$ is written as~\cite{ThompsonPlimpton2009}
\begin{equation}
\label{def_pxy}
p_{xy} = \frac{1}{V} \left[ m \sum_{i} v_{i,x} v_{i,y} - \frac{1}{2} \sum_{i} \sum_{j \neq i} \frac{r_{ij,x} r_{ij,y} \phi' (r_{ij})}{r_{ij}} \right],
\end{equation}
where $\phi(r)$ is the interaction pair potential, $m$ is the mass of a fluid particle, $v_{i,\alpha}$ is the $\alpha$-component of the velocity vector of the $i$th particle, and $r_{ij}$ and $r_{ij,\alpha}$ are the interatomic distance and the $\alpha$-component of the displacement vector between particles $i$ and $j$, respectively.
The first and second terms in the square brackets correspond to the kinetic and potential components (i.e., $p_{xy}^\mathrm{K}$ and $p_{xy}^\mathrm{P}$), respectively, see Eq.~\eqref{decomp_p}.

The three contributors to the SACF $C(t)$, denoted by $C^\mathrm{KK}(t)$, $C^\mathrm{PP}(t)$, and $C^\mathrm{KP}(t)$ (see Eqs.~\eqref{GK_viscosity} and \eqref{deomp_SACF}), are correspondingly defined as
\begin{equation}
C^\mathrm{\circ\star}(t) = \frac{V}{k_\mathrm{B}T}\langle p_{xy}^\circ(0)p_{xy}^\star(t)\rangle,
\end{equation}
where $\circ$ and $\star$ are either K or P.
Note that the time-correlation functions are multiplied by a prefactor of $V / {k_\textnormal{B} T}$ to normalize with respect to the inherent system-size dependence of the microscopic stress tensor.
Then we represent the shear viscosity $\eta$ by the following decomposition:
\begin{equation}
\label{decomp_eta}
\eta = \eta^\mathrm{KK} + \eta^\mathrm{PP} + 2\eta^\mathrm{KP}
\mbox{,  where  }
\eta^\mathrm{\circ\star} = \int_0^\infty C^\mathrm{\circ\star}(t) dt.
\end{equation}

\subsection{MD Simulation}

As a microscopic model of a simple fluid, we consider fluid particles moving in a three-dimensional cubic domain of side length $L$ with periodic boundary conditions.
The interaction between particles is given by the WCA potential:
\begin{equation}
\phi(r) = \left\{ \begin{array}{cc} 4\varepsilon \left[ {\left( \frac{\sigma}{r} \right) }^{12} - {\left( \frac{\sigma}{r} \right) }^{6} + \frac{1}{4} \right], & r \leq 2^{1/6} \sigma , \\ 0, & r > 2^{1/6} \sigma . \end{array} \right.
\end{equation}
We use reduced units of mass, length, and energy, i.e., $m = \sigma = \varepsilon = 1$, along with the Boltzmann constant $k_\textnormal{B} = 1$.
The cutoff radius of the potential is set as $r_\mathrm{c} = 2^{1/6}$.
MD simulations of $N=128,\ldots{},8192$ particles were performed at number densities $n=0.45$, 0.65, and 0.85 and temperature $T=1$.
Under this ambient temperature condition, the three number densities describe typical fluid states; the high density $n=0.85$ corresponds to a dense fluid near the fluid-solid transition point~\cite{IsobeAlder2010}, whereas the low density $n=0.45$ corresponds to a fluid with strong hydrodynamic character.
The simulation box size $L$ is accordingly determined as $L=(N/n)^{1/3}$.
Simulations were conducted using the velocity Verlet algorithm implemented in LAMMPS~\cite{Plimpton1995} with timestep size $\Delta t = 0.002$.
Each equilibrium sample was obtained through equilibration period $\mathcal{T}_\mathrm{equil} = 10^5 \Delta t = 200$ with subsequent production run for period $\mathcal{T} = 10^6 \Delta t = 2000$.
For each set of simulation parameters, a total of $\mathcal{N}=524288$ sample trajectories were calculated.

\begin{figure}
\includegraphics[angle=0, width=0.99\textwidth]{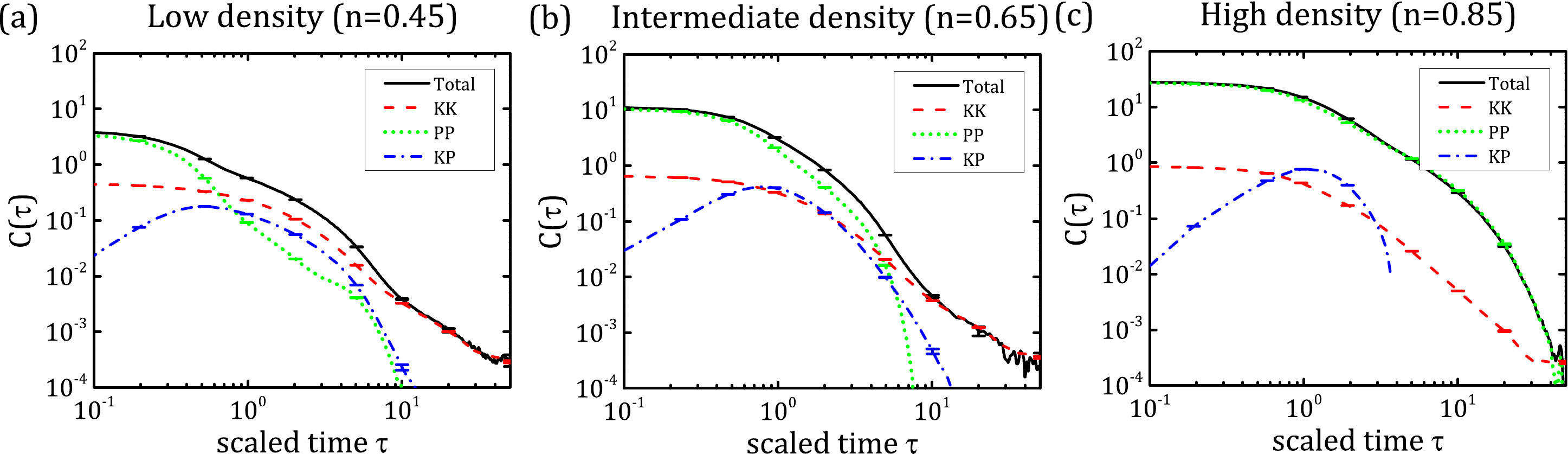}
\caption{\label{fig:sep_sacf_ln}
Three components that make up the normalized SACF $C(\tau)$ at number densities (a) $n=0.45$, (b) $n=0.65$, and (c) $n=0.85$.
The $x$-axis is scaled with the average collision time $t_0$ defined in Eq.~\eqref{def_avg_col_time}.
Results obtained from the largest systems with $N=8192$ are shown and error bars correspond to two standard deviations.
In panel~(c), only positive values of $C^\mathrm{KP}(t)$ are shown.}
\end{figure}

We computed the SACF $C(t)$ and the shear viscosity $\eta$ as well as their components $C^{\circ\star}(t)$ and $\eta^{\circ\star}$ as follows.
From each sample trajectory, instantaneous values of $p_{xy}^\mathrm{K}$ and $p_{xy}^\mathrm{P}$ (see Eq.~\eqref{def_pxy}) were collected at every five timesteps to calculate $C^{\circ\star}(t)$ using a standard time-averaging procedure~\cite{KimBorodin2015}.
The results at different densities are shown in Figures~\ref{fig:sep_sacf} and \ref{fig:sep_sacf_ln}, which will be discussed in Section~\ref{sec_res_sacf}.
We use a scaled time $\tau = t/t_0$ with $t_0$ being the average collision time 
\begin{equation}
\label{def_avg_col_time}
t_0 = \left[ 4n\sigma^2 \left( \frac{\pi k_\mathrm{B}T}{m} \right)^{1/2} g(\sigma) \right]^{-1},
\end{equation}
where $\sigma$ is defined as the position of the first maximum of the radial distribution function $g(r)$.
Numerical time-integration of the correlation functions, $C(t)$ and $C^{\circ\star}(t)$,  was conducted using the trapezoidal rule to obtain the time interals up to time $t$, e.g., $\eta(t) = \int_0^t C(t') dt'$.
For each set of simulation parameters, the $t^*$ value that satisfies $\eta(t^*) \approx \eta = \lim_{t\rightarrow\infty} \eta(t)$ was chosen a posteriori from the time profile of $\eta(t)$ and was used to determine the values of $\eta$ and $\eta^{\star\circ}$.
For each physical quantity, we computed its sample mean and standard deviation over $\mathcal{N}=524288$ samples.

\section{\label{sec_res_sacf}Shear-stress Autocorrelation Function}

We begin with an overall description of the time profiles of the SACF and its components at low, intermediate, and high densities displayed in Figure~\ref{fig:sep_sacf_ln}.
The short-time behavior of the SACF is dominated by the potential term regardless of number density, whereas its long-time behavior is influenced by both kinetic and potential terms resulting in distinctive density-dependent decay patterns.
The slow algebraic decays observed at low and intermediate densities are mainly due to the kinetic correlation $C^\mathrm{KK}(\tau)$, whereas the potential correlation $C^\mathrm{PP}(\tau)$ is responsible for the faster non-algebraic decay at high density.
The cross correlation $C^\mathrm{KP}(\tau)$ is as influential as the other two components at intermediate timescales, but displays negligible impact on the asymptotic long-time behavior of the SACF.

The decay patterns of the kinetic and potential correlations exhibit distinct density dependencies. 
At all three densities, the slope in the log-log plot of $C^\mathrm{KK}(\tau)$ at large $\tau\gtrsim 10$ is insensitive to change in number density, suggesting that the exponent $\alpha$ of the inverse-power decay $C^\mathrm{KK}(\tau)\sim\tau^{-\alpha}$ remains the same as $\alpha\approx 1.5$.
The existence of such a universal exponent is strong evidence that this relaxation process has a hydrodynamic origin~\cite{ErnstHauge1971}.
On the other hand, the potential correlation decays overall faster than algebraically, implying that it has a different physical origin.  
With increasing number density, $C^\mathrm{PP}(\tau)$ increases, leading to an exponential tail at the high density limit.
Our previous MD study on the shear viscosity at high density~\cite{KimHan2018} suggests that the relaxation process involved in the potential correlation has a configurational origin. 

To understand the intrinsic nature of kinetic and potential correlations, in the following sections we systematically examine their density dependence along with the system-size dependence.
In Section~\ref{subsec_res_sacf_a}, we introduce an analytic expression of $C^\mathrm{KK}(t)$ based on molecular hydrodynamics to explain the long-time algebraic decay of the kinetic correlation and examine its validity by comparing with MD simulation results.
In Section~\ref{subsec_res_sacf_b}, we examine the configurational nature of the potential correlation.

\subsection{\label{subsec_res_sacf_a}Hydrodynamic Nature of Kinetic Contribution}

From the mode-coupling theory with the linearized Navier--Stokes (LNS) equations~\cite{ErnstHauge1971, ErpenbeckWood1982}, an analytic expression for the long-time behavior of $C^{\mathrm{KK}}(t)$ in finite systems is given as
\begin{equation}
\label{eq:Ernst_kSum}
C^{\mathrm{KK}}(t) = \frac{2k_{\mathrm{B}}T}{5nL^3} \sum_{\boldsymbol{k} \neq \boldsymbol{0}} \exp \left( - \frac{2 k^2 \eta}{n} t \right),
\end{equation}
where $k$ denotes the magnitude of wavevector $\boldsymbol{k}=\frac{2\pi}{L} (n_x, n_y, n_z)$ with integers $n_x$, $n_y$, $n_z$.
In the thermodynamic limit $L \rightarrow \infty$, the summation of exponential decay functions over wavevectors becomes an inverse-power decay:
\begin{equation}
\label{eq:Ernst_theromdynamicLimit}
C^{\mathrm{KK}}(t) = \frac{2k_{\mathrm{B}}T}{5n} \left( \frac{8 \pi \eta}{n} t \right)^{-3/2}.
\end{equation}
Since this approach assumes that the motion of a molecular fluid is well described by the LNS equations, the validity of Eqs.~\eqref{eq:Ernst_kSum} and \eqref{eq:Ernst_theromdynamicLimit} is not guaranteed for short time $t$ comparable to or shorter than the average collision time $t_0$.
However, our previous MD study on molecular hydrodynamics~\cite{HanKimTalknerKarniadakisLee2018} showed that the dynamics of a molecular fluid is fairly well described even for several collision times. 
Longitudinal velocity field contributions are neglected and thus the effect of sound wave propagation under periodic boundary conditions is not included in Eq.~\eqref{eq:Ernst_kSum}. 

\begin{figure}
\includegraphics[angle=0, width=1.0\textwidth]{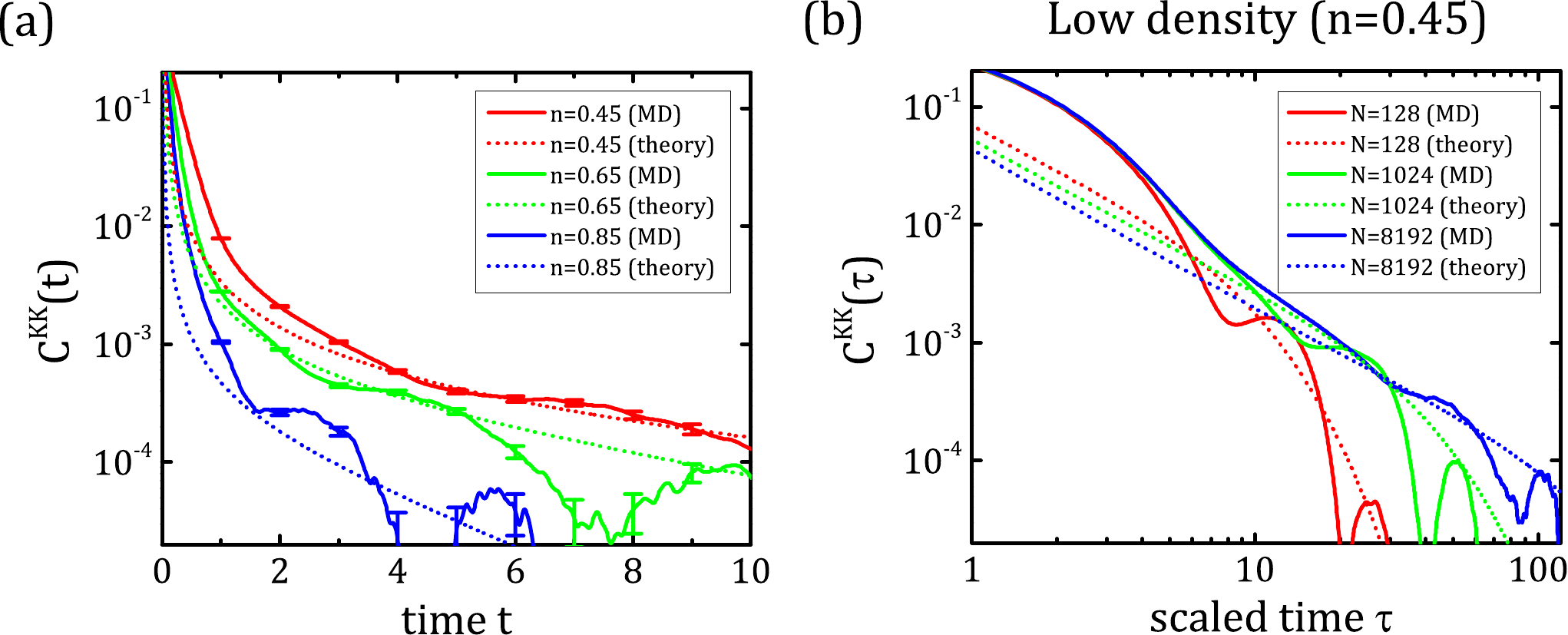}
\caption{\label{fig:sacf_kk}
Comparison of the time profiles of $C^{\mathrm{KK}}(t)$ obtained from MD simulations with those predicted by theory using Eq.~\eqref{eq:Ernst_kSum}.
In panel~(a), results of three densities $n=0.45$, 0.65, and 0.85 are shown for $N=8192$ systems.
In panel~(b), results of three system sizes $N=128$, 1024, and 8192 are shown for low density $n=0.45$.
The $x$-axis is scaled with the average collision time $t_0$ in panel~(b).
Error bars correspond to two standard deviations.}
\end{figure}

\begin{figure}
\includegraphics[angle=0, width=0.40\textwidth]{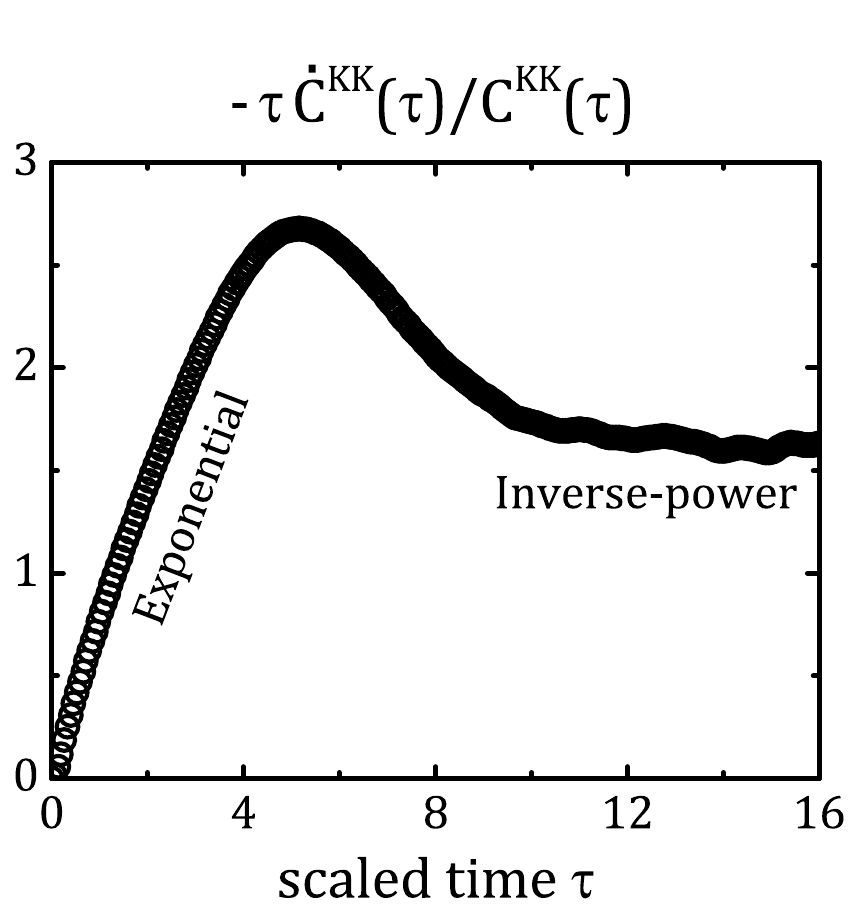}
\caption{\label{fig:sacf_kk_transition}
Time profile of $- \tau \dot{C}^{\mathrm{KK}}(\tau) / C^{\mathrm{KK}}(\tau)$ computed from the $N=8192$ system at $n=0.45$.
For the inverse-power decay $C^\mathrm{KK}(\tau)\sim\tau^{-\alpha}$, the expression gives the exponent $\alpha$.
For the exponential decay $C^\mathrm{KK}(\tau)\sim e^{-\lambda\tau}$, it becomes $\lambda\tau$ (i.e. linear growth in time).}
\end{figure}

In Figure~\ref{fig:sacf_kk}, we first compare the time profiles of $C^{\mathrm{KK}}(t)$ obtained from MD simulations with the theoretical prediction given in Eq.~\eqref{eq:Ernst_kSum}.
For large $t$, LNS predicts the asymptotic behavior of $C^{\mathrm{KK}}(t)$ regardless of number density and system size.
As expected, the oscillations observed in MD results, which are caused by sound wave propagation across periodic boundaries~\cite{ChoiHan2017, HanKimTalknerKarniadakisLee2018}, are not reproduced by the theoretical prediction.
In addition, the discrepancy between MD and LNS is evident at short times, where dynamics is governed by detailed molecular interactions rather than by hydrodynamic laws.
As observed in Figure~\ref{fig:sep_sacf_ln}, while the increase in number density weakens the kinetic correlation, it does not alter the inverse-power decay form of the long-time decay.
It is also observed in Figure~\ref{fig:sacf_kk}b that the time region exhibiting this characteristic decay gradually extends as the system size increases.
Figure~\ref{fig:sacf_kk_transition} clearly shows the transition from the short-time exponential decay to the long-time inverse-power decay observed from the $N=8192$ system at low density $n=0.45$.

\begin{figure}
\includegraphics[angle=0, width=1.0\textwidth]{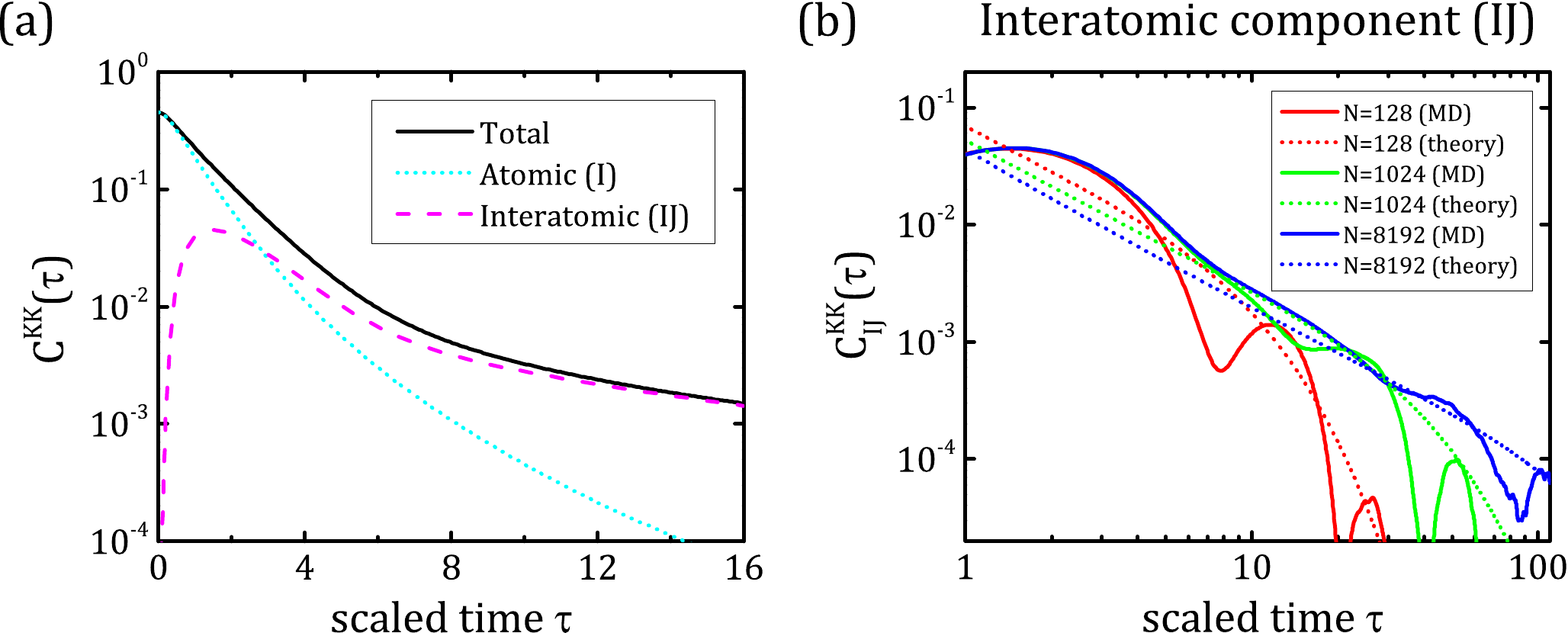}
\caption{\label{fig:sacf_kk_separation}
Decomposition of the kinetic correlation $C^\mathrm{KK}(\tau)$ at low density $n=0.45$.
In panel~(a), the self correlation $C^\mathrm{KK}_\mathrm{I}(\tau)$ and the interatomic correlation $C^\mathrm{KK}_\mathrm{IJ}(\tau)$ are plotted with the total kinetic correlation $C^\mathrm{KK}(\tau)=C^\mathrm{KK}_\mathrm{I}(\tau)+C^\mathrm{KK}_\mathrm{IJ}(\tau)$.
MD results from the $N=8192$ system are used.
In panel~(b), the time profiles of $C_\mathrm{IJ}^{\mathrm{KK}}(\tau)$ obtained from MD simulations with various system sizes are compared with the theoretical prediction from Eq.~\eqref{eq:Ernst_kSum}, cf.\ Figure~\ref{fig:sacf_kk}b.
The $x$-axis is scaled with the average collision time $t_0$.}
\end{figure}

To investigate further the hydrodynamic origin of the kinetic correlation, we subdivide $C^\mathrm{KK}(t)$ into the atomic (or self) correlation $C^\mathrm{KK}_\mathrm{I}(t)$ and the interatomic correlation $C^\mathrm{KK}_\mathrm{IJ}(t)$:
\begin{equation}
\begin{split}
\label{eq:separation_KK}
C^\mathrm{KK}_\mathrm{I}(t) & = \frac{m^2}{Vk_\mathrm{B}T} \bigg\langle \sum_{i} v_{i,x}(0) v_{i,y}(0) v_{i,x}(t) v_{i,y}(t) \bigg\rangle, \\
C^\mathrm{KK}_\mathrm{IJ}(t) & = \frac{m^2}{Vk_\mathrm{B}T} \bigg\langle \sum_i \sum_{j \neq i} v_{i,x}(0) v_{i,y}(0) v_{j,x}(t) v_{j,y}(t) \bigg\rangle.
\end{split}
\end{equation}
Figure~\ref{fig:sacf_kk_separation}a shows the temporal changes in the contributions of these correlations to the total kinetic correlation.
While $C_\mathrm{I}^{\mathrm{KK}}(t)$ dominates at short times,  $C_\mathrm{IJ}^{\mathrm{KK}}(t)$ persists at long times.
Since any term in $C^{\mathrm{KK}}$ at $t=0$ that is odd in velocity component vanishes due to symmetry, $C_\mathrm{IJ}^{\mathrm{KK}}(0)=0$ and thus $C^{\mathrm{KK}}(0) = C_\mathrm{I}^{\mathrm{KK}}(0)$.
However, $C_\mathrm{IJ}^{\mathrm{KK}}(t)$ soon outgrows $C_\mathrm{I}^{\mathrm{KK}}(t)$ as the interatomic correlation starts to have nonzero values.
This transition corresponds to the physical picture that momentum carried by a fluid particle is dissipated to the surrounding particles over time.
In fact, as shown in Figure~\ref{fig:sacf_kk_separation}b, the long-time algebraic tail of $C^\mathrm{KK}(\tau)$ is well described solely by the interatomic correlation  $C_\mathrm{IJ}^{\mathrm{KK}}(\tau)$.

\subsection{\label{subsec_res_sacf_b}Configurational Nature of Potential Contribution}

\begin{figure}
\includegraphics[angle=0, width=1.0\textwidth]{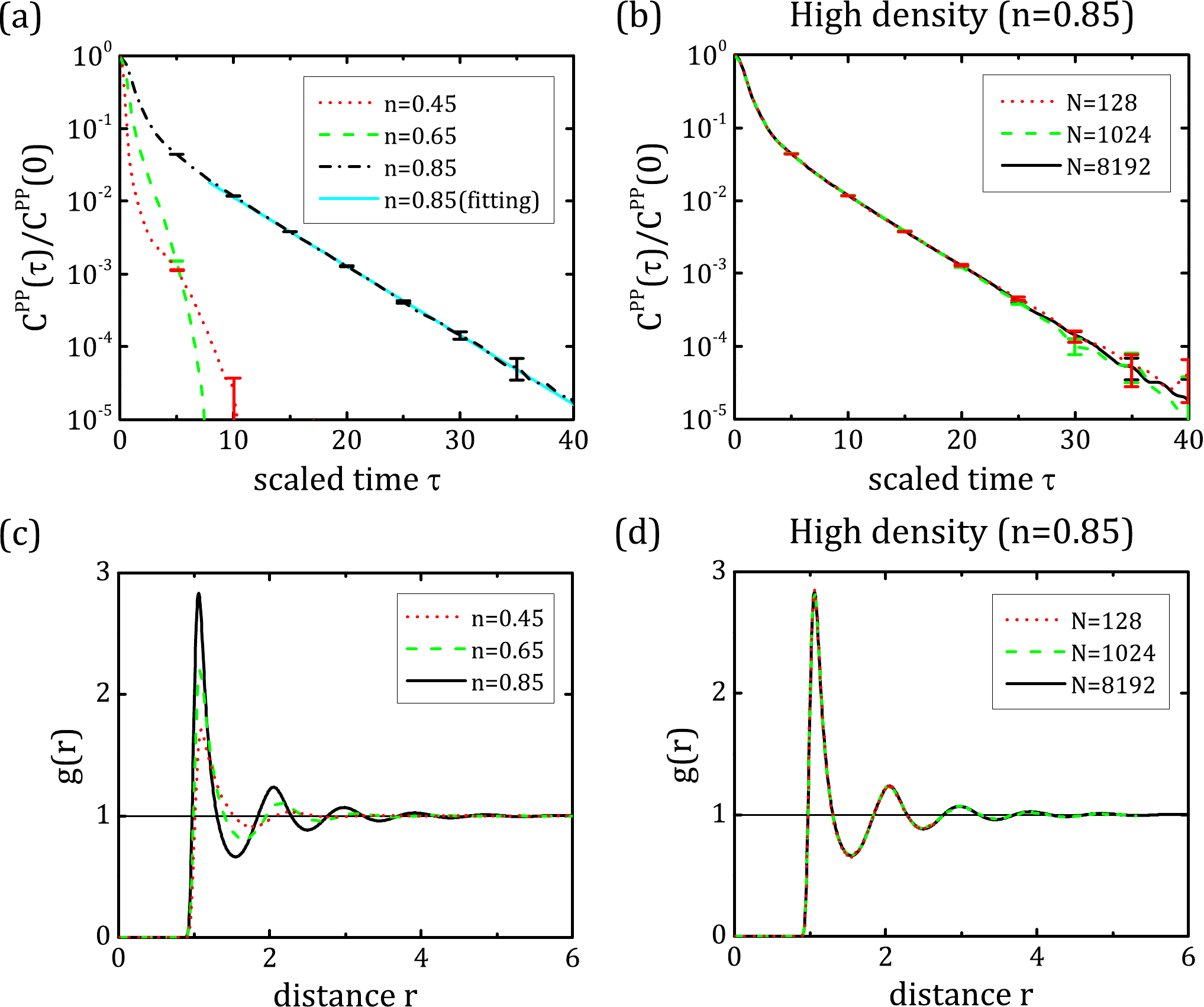}
\caption{\label{fig:sacf_pp}
Normalized potential correlation $C^{\mathrm{PP}}(\tau)/C^{\mathrm{PP}}(0)$ and the radial distribution function $g(r)$.
Panels~(a) and (c) display $C^{\mathrm{PP}}(\tau)/C^{\mathrm{PP}}(0)$ and $g(r)$, respectively, for $N=8192$ systems at different number densities $n = 0.45$, 0.65, and 0.85.
The solid cyan line in panel (a) denotes the exponential fit to MD data at $n=0.85$.
Panels~(b) and (d) display $C^{\mathrm{PP}}(\tau)/C^{\mathrm{PP}}(0)$ and $g(r)$, respectively, at $n=0.85$  for different system sizes $N = 128$, 1024, and 8192.
In panels~(a) and (b), the $x$-axes are scaled with the average collision time $t_0$ and error bars correspond to two standard deviations.}
\end{figure}

Here we study the configurational aspect of the potential-potential component of the SACF.
We first observe the influence of number density on $C^{\mathrm{PP}} (\tau)$.
Figures~\ref{fig:sacf_pp}a shows that the increase in number density strengthens the potential correlation.
At low and intermediate densities, $C^{\mathrm{PP}}(\tau)$ is observed to decay rapidly.
However, owing to numerical imprecision, their functional forms cannot be determined.
On the other hand, an exponential tail is clearly observed at high density $n=0.85$.
The decay pattern appearing after $\tau\approx 5 t_0$ is well-described by an exponential decay with exponent 0.218.
We believe that the different decay patterns of dilute and dense fluids are caused by the difference in the timescales of structural relaxation.
The significance of structural relaxation in $C^{\mathrm{PP}}(t)$ has been repeatedly noted by Isobe and Alder~\cite{IsobeAlder2009,IsobeAlder2010,IsobeAlder2012}.
For dense two- and three-dimensional simple fluids, they have demonstrated the existence of the molasses regime in SACF and related its development to the reorganization of atomic clusters.
The higher the number density, the longer the time it takes for the atomic cluster to dissociate, thereby result in an extensive molasses regime near the fluid-solid transition point.

Figure~\ref{fig:sacf_pp}c compares the radial distribution functions (RDFs) at low, intermediate, and high densities.
As a static equilibrium quantity, the RDF does not provide complete information to compute the potential correlation $C^\mathrm{PP}(t)$ except for the value at $t=0$~\cite{ZwanzigMountain1965},
\begin{equation}
\label{eq:separation_t=0_PP_reduced}
    C^{\mathrm{PP}} (0) = \frac{2 \pi n^2}{15} \int^\infty_0 r^4 \left[ \phi'' (r) + \frac{4}{r} \phi' (r) \right] g(r) dr.
\end{equation}
However, from the intensity of peaks in the RDF, a much slower structural relaxation is expected for high density than for the case of low density.

We next discuss the finite system-size effect on $C^{\mathrm{PP}} (\tau)$.
The influence of system size is not noticeable in Figure~\ref{fig:sacf_pp}b (note, however, that the log scale is used for the values of $C^{\mathrm{PP}} (\tau)$ here).
Contrary to the kinetic correlation (see Figure~\ref{fig:sacf_kk}b), the pattern of the tail is not sensitive to system size and a change in system size does not alter the dynamic feature of $C^{\mathrm{PP}} (t)$.
However, perceptible system-size dependency is observed in the values of $C^{\mathrm{PP}}$ at $t=0$ (see Figure~\ref{fig:vis_pp}a).
In fact, similar size dependency is observed consistently at different times, leading to similar size dependency in $\eta^\mathrm{PP}$ to be discussed in Section~\ref{subsec_res_vis_b}.
At high density, $C^{\mathrm{PP}} (0)$ values show an oscillatory behavior with respect to system size, where increase in system size dampens the magnitude of the oscillation.
At low density, $C^{\mathrm{PP}} (0)$ monotonically increases with increasing system size but at a very slow rate.

In principle, $C^{\mathrm{PP}} (0)$ can be computed using information on the equilibrium distribution such as the RDF, see Eq.~\eqref{eq:separation_t=0_PP_reduced}.
Figure~\ref{fig:sacf_pp}d compares the RDFs obtained from various system sizes.
Essentially the same $g(r)$ is reproduced as for small systems for the range of $r\lesssim L$.
However, peaks appearing beyond this range are missing in the $g(r)$ for small systems.
A more pronounced system-size dependence is expected to be observed at higher densities until all peaks in $g(r)$ are reproduced for sufficiently large systems.
This induces a system-size dependence on $C^\mathrm{PP}(0)$.

\section{\label{sec_res_vis}Shear Viscosity}

\begin{figure}
\includegraphics[angle=0, width=1.0\textwidth]{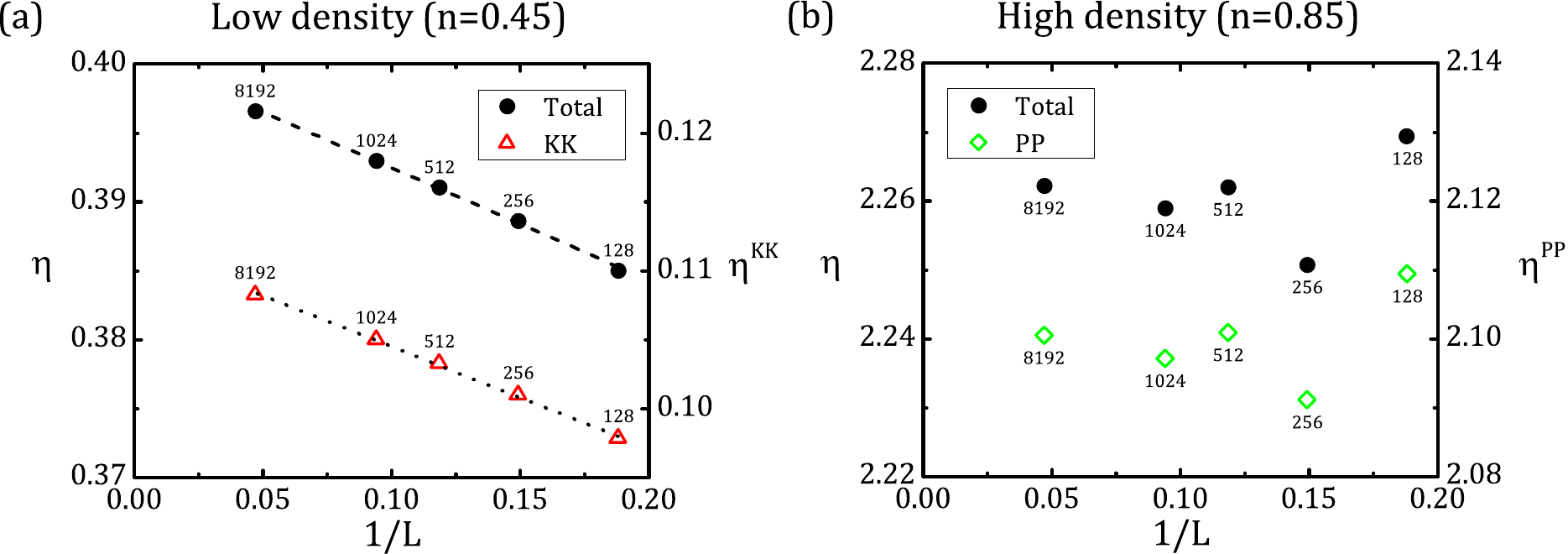}
\caption{\label{fig:vis_sizeEffect}
Finite system-size effects on the shear viscosity and its components.
In panel~(a), $\eta$ and $\eta^\mathrm{KK}$ are compared for low density $n=0.45$.
The dashed and dotted lines denote linear fits to the shear viscosity data and kinetic-kinetic component data, respectively.
In panel~(b), $\eta$ and $\eta^\mathrm{PP}$ are compared for high density $n=0.85$.
Labelled numbers indicate the number $N$ of atoms in the simulation box and standard errors are smaller than the size of symbols.}
\end{figure}

Depending on the number density of the simulation system, the shear viscosity exhibits two types of system-size behavior that are different in nature.
Figure~\ref{fig:vis_sizeEffect} compares the system-size effects on shear viscosity and its components at low and high densities.
While an $L^{-1}$ correction with respect to system size $L$ is observed at low density, shear viscosity at high density shows an oscillatory behavior that dampens with increasing system size.
The latter observation is very similar to that of our previous MD study of the Lennard-Jones fluid near the triple point~\cite{KimHan2018}.
The shear viscosity at intermediate density overall exhibits an $L^{-1}$ scaling behavior but with minor deviations at small simulation systems (not shown).

The effect of the system size on the shear viscosity is determined by the competition between kinetic and potential contributions in the SACF.
At low density, where the portion of the kinetic contribution is significant (as in Figure~\ref{fig:sep_sacf_ln}a), the shear viscosity follows the overall behavior of the kinetic-kinetic component, denoted as the \textit{hydrodynamic} system-size effect, resulting in an $L^{-1}$ scaling behavior.
On the other hand, the SACF of a dense fluid is governed by the potential-potential component (as in Figure~\ref{fig:sep_sacf_ln}c).
Consequently, the shear viscosity is mainly influenced by the system-size dependence of the potential-potential component, denoted as the \textit{configurational} system-size effect.
The difference between the viscosity values of smaller systems and the largest system are smaller than 3 percent for the low density and 1 percent for the high density.
This is consistent with the weak system-size dependence observed in previous MD studies~\cite{DaivisEvans1995, YehHummer2004, MoultosZhangTsimpanogiannisEconomouMaginn2016}.

In the following subsections, we study the hydrodynamic and configurational system-size effects on shear viscosity based on the observations made in Section~\ref{sec_res_sacf}.
Empirical formulas that describe the scaling behavior of the hydrodynamic system-size effect are proposed in Section~\ref{subsec_res_vis_a}.
The relation between the complex size-dependent behavior of shear viscosity and that of $C^{\mathrm{PP}}(t=0)$ value is discussed in Section~\ref{subsec_res_vis_b}.

\subsection{\label{subsec_res_vis_a}Hydrodynamic System-Size Effect}

In Section~\ref{subsec_res_sacf_a}, we separated $C^\mathrm{KK}(t)$ into the atomic contribution $C_\mathrm{I}^\mathrm{KK}(t)$ and the interatomic contribution $C_\mathrm{IJ}^\mathrm{KK}(t)$ and verified that the latter is responsible for the hydrodynamic long-time tail in $C^\mathrm{KK}(t)$ while the other dominates at shorter times.
Likewise, we separate $\eta^\mathrm{KK}$ into two components, $\eta_\mathrm{I}^\mathrm{KK}$ and $\eta_\mathrm{IJ}^\mathrm{KK}$, and study their respective system-size dependence.
Based on scaling relations observed in the MD simulation data of $C_\mathrm{I}^\mathrm{KK}(t)$ and $C_\mathrm{IJ}^\mathrm{KK}(t)$, their empirical formulas are proposed.
From the relation $\eta_\star^\mathrm{KK}=\int_0^\infty C^\mathrm{KK}_\star(t)dt$ with $\star=\mathrm{I}$ or $\mathrm{IJ}$, the corresponding finite size effects of $\eta_\star^\mathrm{KK}$ are deduced.

\begin{figure}
\includegraphics[angle=0, width=1.0\textwidth]{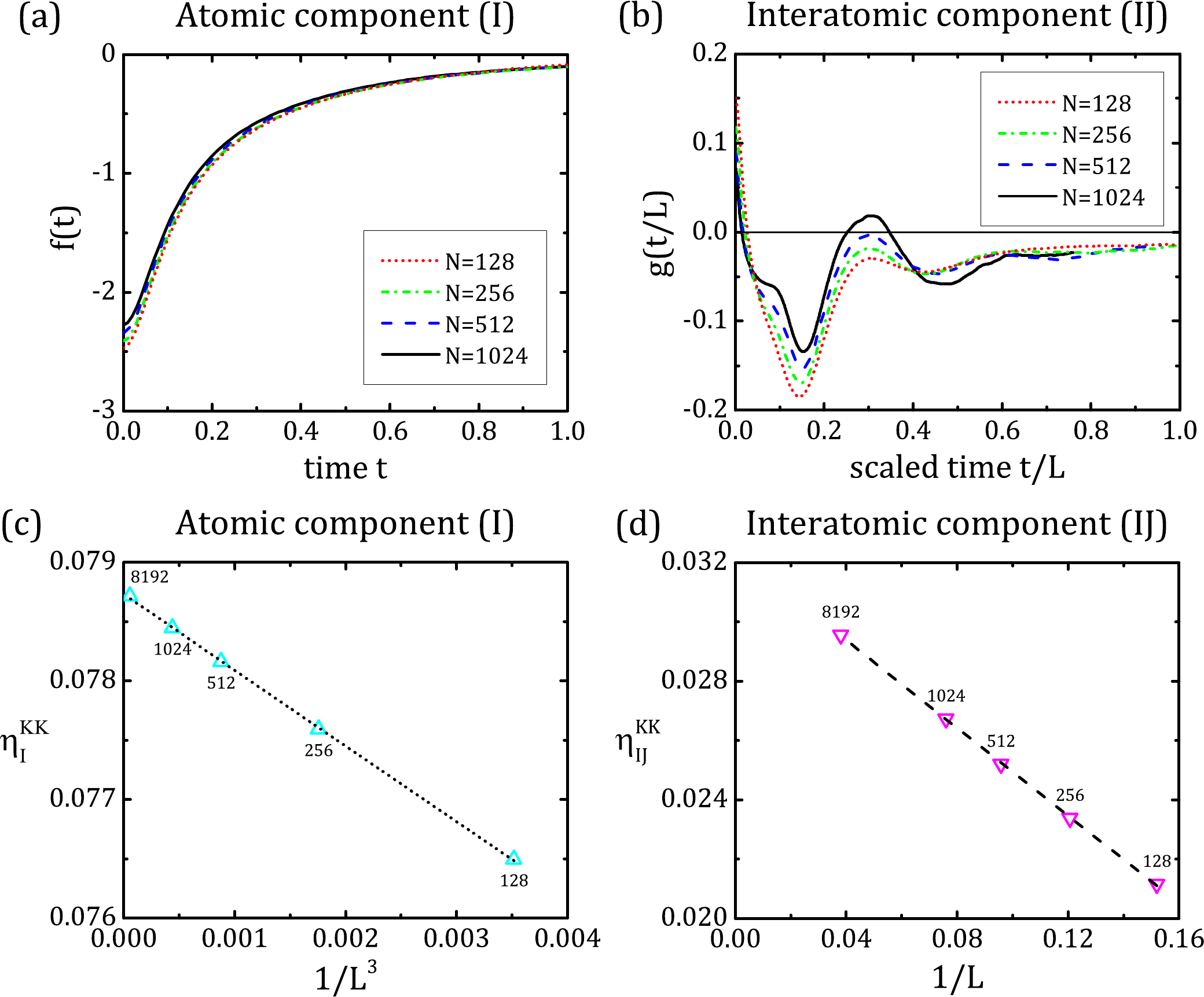}
\caption{\label{fig:vis_kk_scaling}
Scaling behaviors of the atomic and interatomic contributions in $C^{\mathrm{KK}}(t)$ and $\eta^{\mathrm{KK}}$ observed in the low density case $n=0.45$.
In panels~(a) and (b), the time profiles of $f(t)$ and $g(t/L)$ determined from MD data with various system sizes $L$ are shown respectively.
In panels~(c) and (d), the values of $\eta^\mathrm{KK}_\mathrm{I}$ and $\eta^\mathrm{KK}_\mathrm{IJ}$ are plotted versus $1/L^3$ and $1/L$, respectively.
Linear fits to the data are also shown.
Labelled numbers indicate the number of atoms in the simulation box.
Standard errors are smaller than the size of symbols.}
\end{figure}

In Figures~\ref{fig:vis_kk_scaling}a and \ref{fig:vis_kk_scaling}b, we confirm that $C_{\mathrm{I},L}^{\mathrm{KK}}(t)$ and $C_{\mathrm{IJ},L}^{\mathrm{KK}}(t)$ for finite system size $L$ satisfy the following asymptotic formulas: 
\begin{align}
\label{eq:scaling_sacf_i}
C_{\mathrm{I},L}^{\mathrm{KK}}(t) &= C_{\mathrm{I},\infty}^{\mathrm{KK}}(t) + L^{-3} f(t), \\
\label{eq:scaling_sacf_ij}
C_{\mathrm{IJ},L}^{\mathrm{KK}}(t) &= C_{\mathrm{IJ},\infty}^{\mathrm{KK}}(t) + L^{-2} g\left(\frac{t}{L}\right) ,
\end{align}
where $C_{\mathrm{I},\infty}^{\mathrm{KK}}(t)$ and $C_{\mathrm{IJ},\infty}^{\mathrm{KK}}(t)$ denote the corresponding correlations in the thermodynamic limit $L\rightarrow\infty$.
By assuming MD data from the largest $N=8192$ system at $n=0.45$ as bulk data, the time profiles of $f(t)$ and $g(t/L)$ were computed for each small system size using Eqs.~\eqref{eq:scaling_sacf_i} and \eqref{eq:scaling_sacf_ij}.
The remarkable coincidence of the time profiles obtained from various small system sizes indicates that our empirical asymptotic expressions are valid.

By applying the Green--Kubo relation to Eqs.~\eqref{eq:scaling_sacf_i} and \eqref{eq:scaling_sacf_ij}, we obtain the finite system-size correction of $\eta^\mathrm{KK}_L$:
\begin{equation}
\label{eq:scaling_vis}
    \eta_{L}^{\mathrm{KK}} = \eta_{\infty}^{\mathrm{KK}} + \frac{F}{L^3} + \frac{G}{L} ,
\end{equation}
where $F=\int_{0}^{\infty} f(t) dt$ and $G=\int_{0}^{\infty} g(t) dt$.
Figures~\ref{fig:vis_kk_scaling}c and \ref{fig:vis_kk_scaling}d display the actual scaling behaviors of $\eta_\mathrm{I}^{\mathrm{KK}}$ and $\eta_\mathrm{IJ}^{\mathrm{KK}}$ observed in MD simulations, which are consistent with Eq.~\eqref{eq:scaling_vis}.

Different finite system-size effects on $C_\mathrm{I}^{\mathrm{KK}}(t)$ and $C_\mathrm{IJ}^{\mathrm{KK}}(t)$ lead to different system-size corrections of $\eta^\mathrm{KK}$.
The leading $L^{-1}$ correction is obtained from the interactomic correlation $C_\mathrm{IJ}^{\mathrm{KK}}(t)$.
As discussed in Section~\ref{subsec_res_sacf_a} using Figure~\ref{fig:sacf_kk_separation}, $C_\mathrm{IJ}^{\mathrm{KK}}(t)$ is responsible for the long-time tail of $C^\mathrm{KK}(t)$.
Hence, the main finite system-size effect in $\eta^\mathrm{KK}$ is attributed to the disturbance in the long-time tail.

\subsection{\label{subsec_res_vis_b}Configurational System-Size Effect}

In our previous study on the shear viscosity of a \textit{dense} fluid~\cite{KimHan2018}, we observed the remarkable resemblance between the size-dependent behaviors of $C(0)$ and the shear viscosity $\eta$ and drew attention to the significance of the configurational contribution in the estimation of finite system-size effects on shear viscosity.  
However, we were not certain whether the relation holds at intermediate and low densities and therefore did not claim the applicability of the uncertainty quantification method to other liquid states.
Here we confirm that the system-size dependencies of $C^{\mathrm{PP}}(0)$ and $\eta^{\mathrm{PP}}$ are in consistent agreement \textit{regardless of number density}, thereby giving validity to our quantification method for \emph{configurational} system-size effect.

\begin{figure}
\includegraphics[angle=0, width=1.0\textwidth]{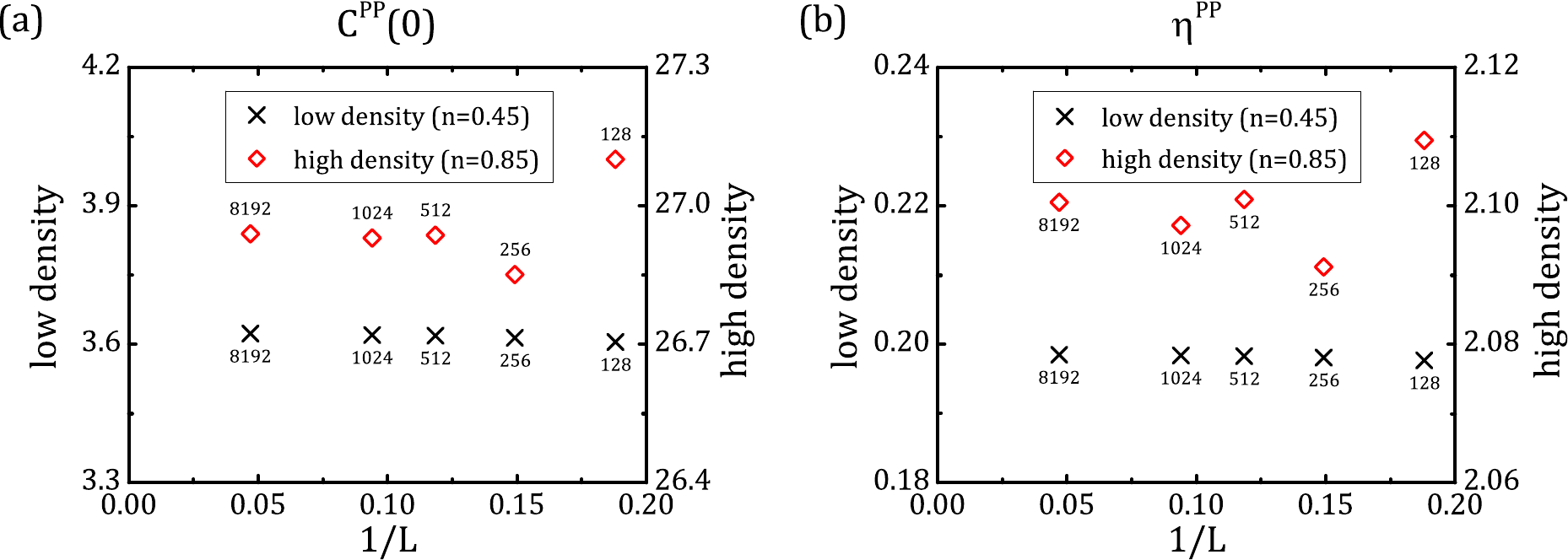}
\caption{\label{fig:vis_pp}
Finite system-size effects on $C^{\mathrm{PP}} (0)$ and $\eta^{\mathrm{PP}}$ at number densities $n=0.45$ and 0.85.
In panel~(a), values of the potential component of the SACF at $t=0$, $C^{\mathrm{PP}} (0)$, are plotted versus the reciprocal of the system size $L$.
In panel~(b), values of the potential component of the shear viscosity, $\eta^\mathrm{PP}$, are accordingly plotted.
Standard errors are smaller than the size of symbols.}
\end{figure}

Figure~\ref{fig:vis_pp} compares the system-size dependencies of $C^{\mathrm{PP}}(0)$ and $\eta^{\mathrm{PP}}$ at $n=0.45$ and 0.85.
While the patterns of each quantity at the two densities are significantly different, the patterns of the two quantities at each density are remarkably similar.
The damped oscillatory behavior of $\eta^{\mathrm{PP}}$ at high density is also observed in $C^{\mathrm{PP}}(0)$.
On the other hand, a weak monotonic increase of $\eta^{\mathrm{PP}}$ at low density with respected to increasing system size is consistent with that of $C^{\mathrm{PP}}(0)$.
In other words, the finite system-size effect of $\eta^{\mathrm{PP}}$ is well reproduced by $C^{\mathrm{PP}}(0)$.

While $C^\mathrm{PP}(t)$ and $\eta^\mathrm{PP}$ are related via the Green-Kubo relation, it is important to note that this does not automatically guarantee the remarkable resemblance of the finite system-size effects on $C^\mathrm{PP}(0)$ and $\eta^\mathrm{PP}$.
In fact, such a coincidence is not expected to happen for the pair of kinetic components, $C^\mathrm{KK}(0)$ and $\eta^\mathrm{KK}$.
In the potential case, the system-size dependence of $C^\mathrm{PP}(t)$ at each time $t$ remains similar resulting in a similar behavior for $\eta^\mathrm{PP}$, whereas in the kinetic case the main finite system-size effect on $\eta^\mathrm{KK}$ results from the change in the long-time tail.  
Hence, while $\eta^\mathrm{PP}$ is a dynamic property, its finite system-size effect can be roughly estimated from a static quantity $C^\mathrm{PP}(0)$, the accurate value of which can be much more easily computed.

\section{\label{sec_conclusion}Conclusions}

The calculation of transport coefficients has long been of interest and importance, see Ref.~\cite{RiceGray1965}.
We have studied the density-dependent finite system-size effects on the shear-stress autocorrelation function $C(t)$ and the shear viscosity coefficient $\eta$ by investigating hydrodynamic and configurational nature in the relaxation process of the off-diagonal shear-stress tensor component $p_{xy}$.
Systematic MD simulations for a three-dimensional simple fluid at low, intermediate, and high densities have revealed that the shear viscosity of a dense fluid exhibits an oscillatory behavior that dampens with increasing system size $L$, whereas that of a dilute fluid has an $L^{-1}$ finite system-size correction.
The former finite system-size effect at high density was identified to originate from configurational nature of the potential-potential component of the SACF, $C^\mathrm{PP}(t)$, whereas the latter at low density was shown to have a hydrodynamic origin arising from the kinetic-kinetic component, $C^{\mathrm{KK}}(t)$.
Competition between these two contributions explains not only the strong density dependence of the shear viscosity but also that of its finite system-size effects.

Using analytic results obtained from the mode-coupling approach with the linearized Navier--Stokes equations~\cite{ErnstHauge1971, ErpenbeckWood1982}, we performed a quantitative analysis on the kinetic correlation $C^{\mathrm{KK}}(t)$ and the kinetic component $\eta^\mathrm{KK}$ of the shear viscosity.
The inverse-power decay of $C^{\mathrm{KK}}(t)$ was clearly observed in MD simulations at low and intermediate densities, which confirms the applicability of molecular hydrodynamic theory~\cite{ChoiHan2017, HanKimTalknerKarniadakisLee2018}.
Using the scaling behavior of the finite system-size effects of $C^{\mathrm{KK}}(t)$, we showed that the $L^{-1}$ correction of $\eta^\mathrm{KK}$ is due to the change in the long-time tail of $C^{\mathrm{KK}}(t)$.
As seen Figure~\ref{fig:sacf_kk}b, the latter change includes an oscillatory disturbance caused by traveling of sound waves across periodic boundaries and a cross-over from the algebraic decay to an exponential regime ascribable to the cut-off at low wavenumbers introduced by periodic boundaries.
Similar behaviors have been observed in the velocity autocorrelation function~\cite{ChoiHan2017, AstaLevesqueVuilleumierRotenberg2017} and  
the same form of correction has been derived for the self-diffusion coefficient~\cite{DunwegKremer1993, YehHummer2004}.
Contrary to the kinetic correlation, an analytic approach to determine the potential correlation $C^\mathrm{PP}(t)$ has not been successful due to the complicated structure of a nonlinear four-particle correlation function which plays a role in the structural relaxation.
Motivated by our previous MD simulation study~\cite{KimHan2018}, we focused on the remarkable resemblance between the finite system-size effects on $C^\mathrm{PP}(0)$ and $\eta^\mathrm{PP}$ regardless of density.
This observation provides a practical suggestion of using $C^\mathrm{PP}(0)$ to roughly estimate the finite system-size effect on $\eta^\mathrm{PP}$.
As a static equilibrium quantity, the accurate value of $C^\mathrm{PP}(0)$ can be much more easily computed than $\eta^\mathrm{PP}$ and thus the critical system size for an accurate value of $\eta^\mathrm{PP}$ can be efficiently estimated.
We also point out that $C^\mathrm{PP}(0)$ can be investigated analytically using the radial distribution function.

\begin{acknowledgments}
This work was supported in part by the U.S.\ Army Research Laboratory and was accomplished under Cooperative Agreement No.\ W911NF-12-2-0023, ``Alliance for the Computationally-guided Design of Energy Efficient Electronic Materials (CDE3M)''.
\end{acknowledgments}

\bibliographystyle{unsrt}
\bibliography{manuscript.bbl}

\end{document}